\documentclass[a4paper]{jpconf}
\usepackage{blindtext}
\usepackage{graphicx}

\usepackage[square,sort&compress]{natbib}
\setcitestyle{numbers}

\begin{document}
\nocite{*}

\title{Probing Neural Networks for the Gamma/Hadron
Separation of the Cherenkov Telescope Array}


\author{Lyard E, Walter R, Sliusar V and Produit N, for the CTA Consortium}

\address{Department of Astronomy, University of Geneva, Switzerland.}

\ead{etienne.lyard@unige.ch}

\begin{abstract}
We compared convolutional neural networks to the classical boosted decision trees for the separation of 
atmospheric particle showers generated by gamma rays from the particle-induced background. 
We conduct the comparison of the two techniques applied to simulated 
observation data from the Cherenkov Telescope Array. We then looked at the Receiver Operating Characteristics (ROC) 
curves produced by the two approaches and discuss the similarities and differences between both. 
We found that neural networks overperformed classical techniques under specific conditions. 

\end{abstract}


\section{Introduction}

Machine learning made spectacular advances during the last few years. Deep convolutional 
neural networks (CNNs) emerged as a very powerful technique thanks to advances in  
algorithms, data availability and overall computational power. CNNs recently proved to be effective also for 
TeV astrophysics \cite{hess} to separate the signal from an astronomical source
from the cosmic-ray background. The Cherenkov Telescope Array (CTA) \cite{cta} will be the next-generation ground-based 
gamma-ray observatory, composed of more than one-hundred telescopes at two
observation sites. Its sensitivity will improve by an order of magnitude compared to
existing facilities. Improving the data analysis techniques to better discriminate the
observed gamma rays from the cosmic rays would allow to better resolve the observed sources and also to 
reduce the observation time needed to obtain enough significance. \\
In this paper we focus on the signal extraction and 
we evaluate the performance of CNNs compared to Boosted Decision Trees (BDTs) \cite{bdt} which are 
commonly used for this task and in particular in the EventDisplay analysis package \cite{eventdisplay}. Other techniques are commonly employed
such as random forests \cite{random_forests} or maximum-likelihood \cite{model}. 
To perform this comparison, we took the Event parameter output of the EventDisplay analysis of simulated 
events from a Monte-Carlo (MC) production of CTA. We picked the dataset that contained 
the most realistic observation conditions and applied CNNs to it. We then compared the 
obtained Signal/Background separation performances with the one from the BDTs from EventDisplay.

\subsection{Cherenkov Astronomy}

Cherenkov Astronomy studies very high-energy $\gamma$-ray emission from the Galaxy 
and beyond. Above a few GeV, the flux from the sources is too small to be 
detected with a compact instrument. Instead,
it relies on the interaction between high-energy gamma rays and the Earth's atmosphere.
These interactions produce showers of particles travelling faster than light in the atmostphere which
thus emit Cherenkov light. This light is detected by arrays of large telescopes and very fast
cameras. Each individual shower detection is called an \emph{event}, which combines images seen in coincidence in several telescopes. \\
Images typically look like elongated ellipses, and these can be combined to give event-level parameters. They can be of several 
types, namely photonic (gammas), hadronic or electronic. It is the gamma events
that are of interest, as hadronic events originate from cosmic-rays.
The discrimination of events cannot be fully accurate and thus gamma events
are discarded hence reducing the sensitivity of the facility. The opposite also occurs, with 
many hadronic events being classified as gammas, hence contaminating the measurement with background noise. 

\subsection{Convolutional Neural Networks}

CNNs extract features from images and combine them to derive higher level knowledge. 
Their architecture has grown more complex over the years, and they now outperform humans in 
recognizing a large variety of items in images. The current best performing 
network, called Squeeze-and-Excitation \cite{squeeze} was able to recognize which of the 1000 
possible object categories appeared on pictures with an accuracy of 97.75 percents during the 
2017 annual ImageNet challenge \cite{ILSVRC15}. 
This approach started to be applied in astrophysics to separate signal from sources from
background noise. \cite{SCHA17} used CNNs and generative adversarial 
networks to recover features in astrophysical images of galaxies beyond the deconvolution 
limit. CNNs were also used by \cite{HEZA17} to deconvolve strongly-lensed 
images of galaxies. \cite{SCHA18} used a similar approach to detect 
such images while \cite{STAR18} used Generative Adversarial Networks to separate quasar point
sources from the light of its host galaxy. \cite{ERDM18} used deep-learning 
to reconstruct air showers from data coming from the Pierre Auger Observatory \cite{auger}. 
Eventually, \cite{DANI18} applied deep learning to gravitational waves 
detection and the estimation of their parameters using Laser Interferometer Gravitational-Wave Observatory data \cite{LIGO}. 

\section{Proposed method}
The performances of neural networks can be quite difficult to evaluate because they are very 
sensitive to the training and validation datasets that are used. The computer vision 
community solved this issue by defining a standard dataset to be used both for training
and evaluation of new architectures \cite{imagenet}. Such dataset does not yet exist in high-energy 
astrophysics. Thus, we decided to evaluate the performances of CNNs with respect to what 
is the current standard in the field, namely BDTs. 
CTA has produced a standard analysis of the simulated data, and it is against this classification
that we evaluated our neural networks. The CTA standard analysis relies on the EventDisplay
package \cite{eventdisplay} that was originally developed for the VERITAS experiment \cite{veritas}. Another package named
MARS \cite{mars} is used to crosscheck the results.  
We used exactly the same datasets for both the BDTs and the neural networks. Both have exactly
the same amount of data to work with, hence we believe that such a comparison is fair and makes
sense in this context. We did not perform the EventDisplay analysis ourselves as the output of 
the analysis is available to all CTA consortium members. 
Both approaches are compared by plotting their Receiver Operating Characteristics (ROC) curves. 
The Area Under Curve (AUC) is used to assess the methods' overall performances, while subtle 
differences between the curves are discussed to make predictions about their true performances. 

\section{Monte-Carlo Data}
We decided to use the datasets from night sky background (NSB) studies 
to perform this comparison. These datasets are well suited because they were designed to represent
the standard operating conditions of CTA. We stick to the standard NSB level as it is the one that
is the closest to the expected nominal conditions and used only medium-size telescopes data, as did the BDT analysis. The CTA expert helped us retrieve a list of 
events that were used to train and validate the BDTs. The datasets contain only diffuse protons and 
diffuse gammas. No electrons were included because it was not the primary goal of this study, but 
also because electrons would be much more difficult to differentiate and give a diffuse background at 
a lower level than typical gamma-ray sources. \\
A preliminary cut performed by the CTA analysis removed the most obvious background events from 
the datasets. As a consequence, the performance curves given in a later section do not take the 
full data into account, but rather only the portion of the \emph{difficult} background events.
Moreover, the simulations focused on gamma-detection efficiency under various NSB conditions. Thus
more signal events were simulated compared to what one can reasonably expect from a real instrument.
Consequently the performance curves given in the results section cannot be used to estimate the 
overall performance of the method, but only its performance with respect to BDTs. 

\subsection{Neural Network Input}
BDTs operate on event parameters extracted from the raw events data. In contrast, our CNN architecture 
operates directly on the raw data. Nevertheless, we applied a data reduction step to make the datasets 
easier to work with, as follow:
\begin{itemize}
\item \emph{Waveform integration}. Most Cherenkov cameras record short movies of up to 300ns in duration, 
with each frame lasting between 0.5 and 4 nanoseconds depending on the instrument. We integrated
the signal of each pixel to reduce the dimensionality. Instead of working with \emph{N} time-samples
for each pixel, we ended up with two values: integrated charge and time-of-maximum. 
\item \emph{Image calibration}. We applied a calibration step to work with photo-electrons rather than 
integrated ADC counts. The time of maximum was kept as an index to the sample.
\item \emph{Image normalization}. For the intensity value, we normalize the image so that the maximum 
pixel value is always 1000. Rather than normalizing to 1, we preferred to remain in the integer domain and 
be able to work with smaller files. 
\end{itemize}

\begin{figure}[h!]
\begin{centering}
\resizebox{0.8\columnwidth}{!}{\includegraphics{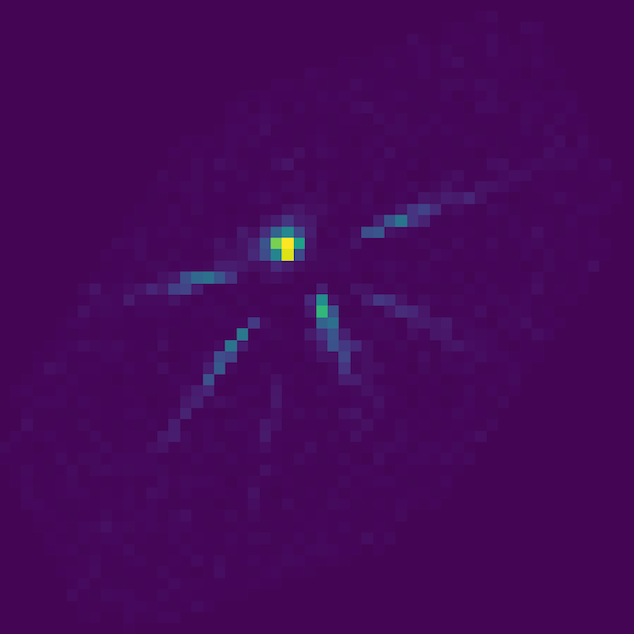}
							   \includegraphics{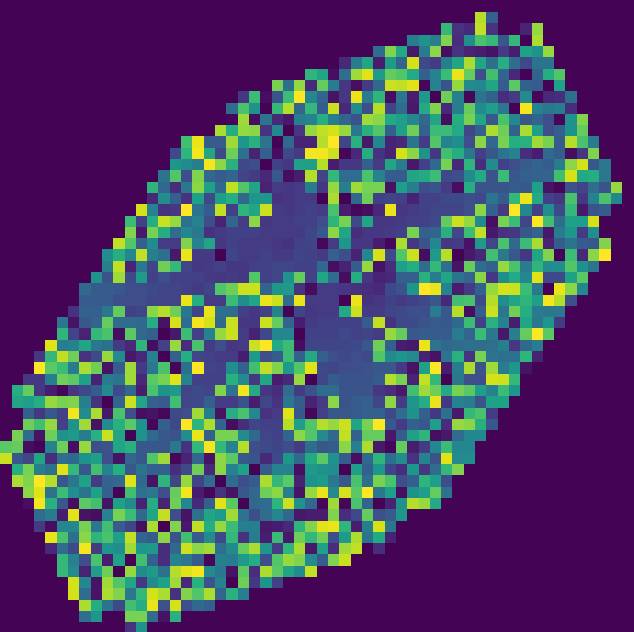}
							   \includegraphics{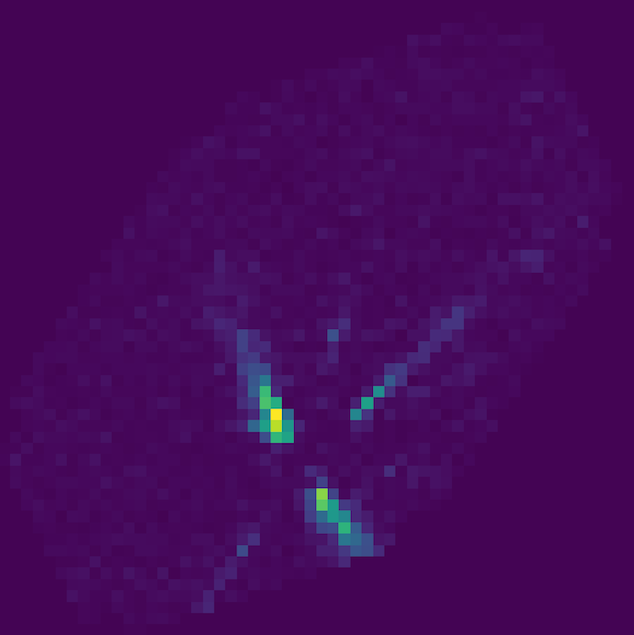}
							   \includegraphics{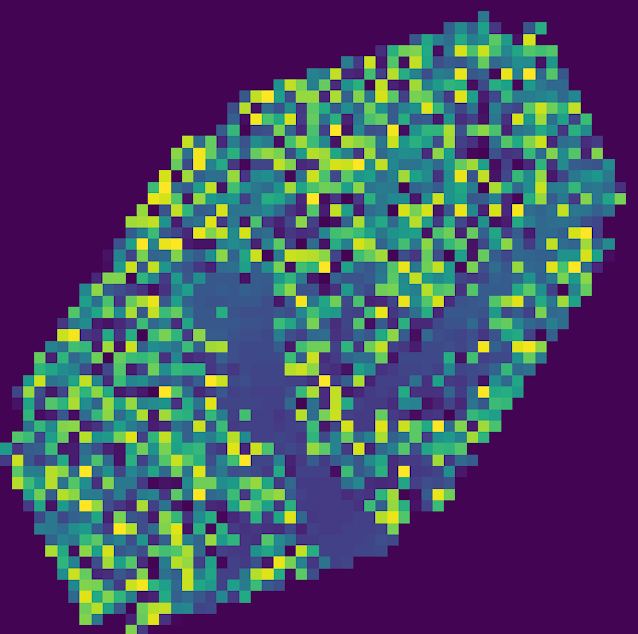}}

\caption{Example of a gamma and hadron events transformed to a square image.
From left to right: gamma intensity, gamma timing, hadron intensity, hadron timing.}
\label{fig:inputgamma}
\end{centering}
\end{figure}

Besides the steps above, existing high-level CNN packages work with square images with multiple 
channels (red, green and blue). On the other hand, our reduced datasets contain
hexagonal images with two values. We applied a geometry conversion step to transform hexagonal 
images into square ones (figure \ref{fig:inputgamma}). This introduces 
a geometrical bias that remains to be addressed. Multi-telescope data was dealt with by simply 
stacking all telescopes' images into a single image, as seen on the example events images. 

\subsection{Energy Bands}
The BDTs that we compared against operated on specific energy bands. This is a common approach meant to 
simplify models and speed up training time. We applied
the same splitting of the data to train the CNNs. 
We tested 5 energy bands, as shown in table \ref{tab:resultsoverview}. Each energy band had
 approximately two 
times more signal than background events. Half the events were used for training, half for 
validation. 

\section{Network Architecture}
Following the survey work from \cite{nieto}, we decided to start from the best network
according to their study, namely InceptionV3 \cite{inception}. We did not explore the hyper-parameters space either and
only used ADADELTA \cite{ADADELTA} for the optimizer and binary
cross-entropy \cite{CROSS} for the loss function. We focused instead on the network architecture, and quickly
found out that InceptionV3 has too many layers for the task at hand. This makes sense
as InceptionV3 was designed to classify images into one thousand categories, while we only have two possible
outputs (gamma / hadron). 

\begin{figure*}
\begin{centering}
\resizebox{0.8\columnwidth}{!}{\includegraphics{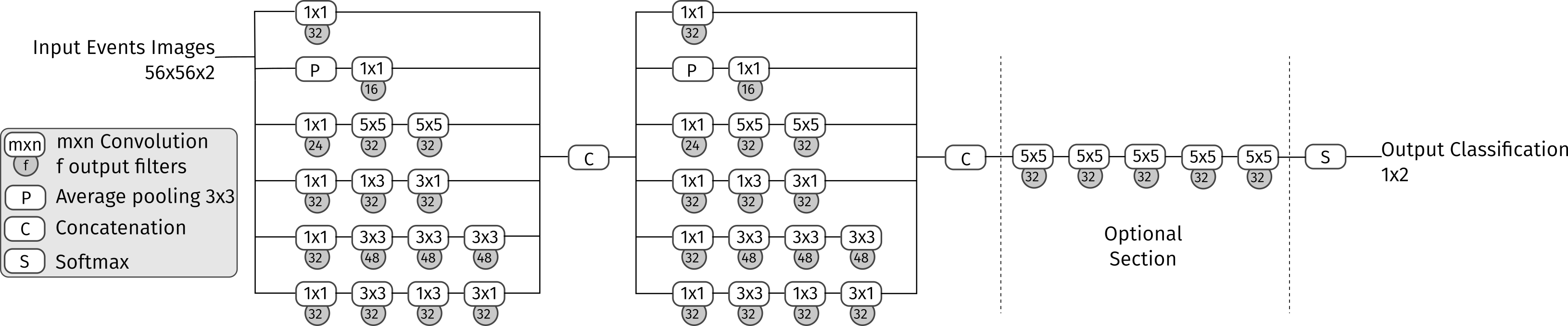}}
\caption{Overall architecture of the CNN used for this study. The optional section is only present
for the model with 595k parameters.}
\label{fig:overallarchitecture}
\end{centering}
\end{figure*}

After some trial-and-error we ended up with a baseline architecture that has nine layers for a 
total of 290k parameters (DL290k - figure \ref{fig:overallarchitecture}). We also tested two variants of this architecture:
\begin{itemize}
\item Simplified: same as the baseline architecture, but with half the number of filters for each convolution kernel. 
Total parameters: 18k (DL18k)
\item Extended: same as the baseline architecture, but with extra convolutions before
the softmax layer \cite{softmax}. Total parameters: 595k (DL595k)
\end{itemize}

There was no dropout layer \cite{dropout} included in the model. The reason for this is two-fold. First, adding
dropout layers significantly decreased the performance of the model and slightly increased the 
training time. Second, because we have virtually unlimited simulated data to train the models, overfitting
can be dealt with by increasing the size of the training dataset.

\section{Results}
\begin{figure*}
\begin{centering}
\resizebox{1.0\columnwidth}{!}{\includegraphics{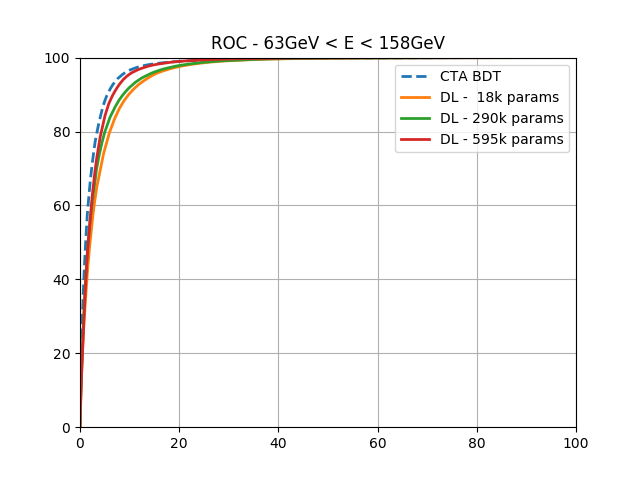}
							   \includegraphics{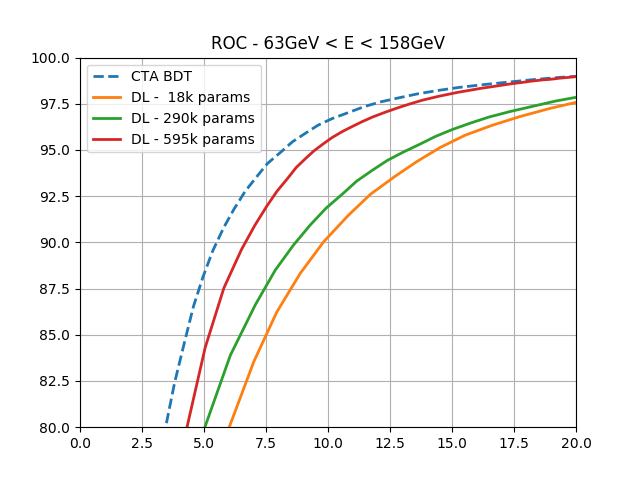}
								\includegraphics{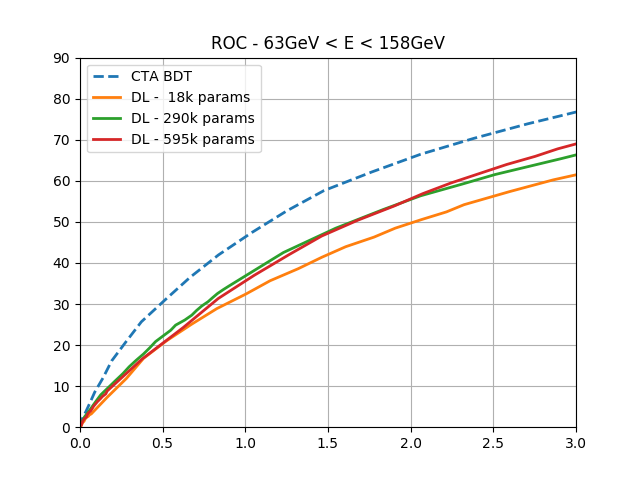}}
\resizebox{1.0\columnwidth}{!}{\includegraphics{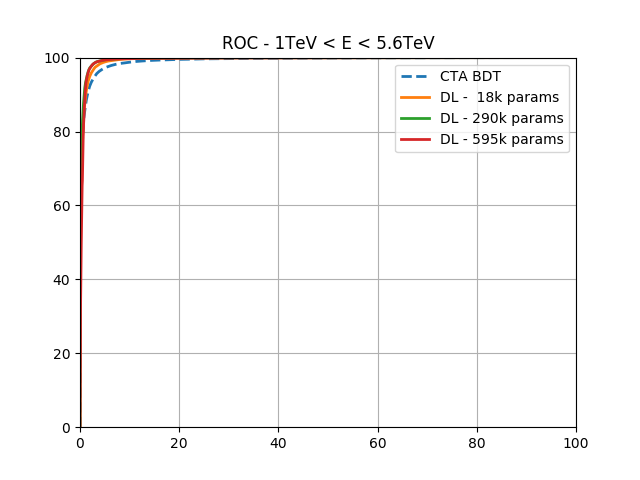}
							   \includegraphics{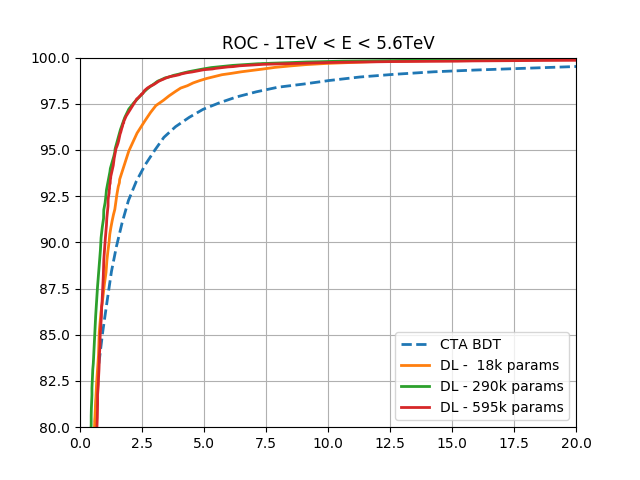}
							   \includegraphics{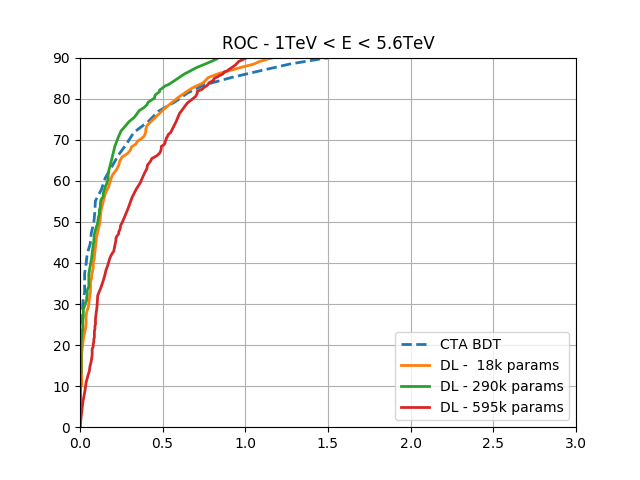}}
\caption{ROC curves for energy from 63 to 158GeV, for which BDTs outperformed CNNs, and from 1 to 5.6TeV, 
for which CNNs outperformed BDTs. Left: zoomed out view, middle: zoom in the central region, right: zoom in the beginning region}
\label{fig:resultsoverall}
\end{centering}
\end{figure*}

Even though results are presented in the form of ROC curves below, in reality the ratio of signal/background
events is in the order of $\frac{1}{10000}$. It is thus very important that the ROC curve be as steep as
possible so as to limit the contamination of the signal by background events.
The CNNs peaked between epoch 7 and 59 depending on the training parameters before starting to overfit on the
training data. They performed in a very similar way to the BDTs, as seen in table \ref{tab:resultsoverview}. 
This result is quite encouraging, as no a-priori knowledge was given to the CNNs. On the contrary a lot of human
expertize was put in the training of the BDTs, even if the models that we compared against may not be the best
possible ones. CNNs outperformed BDTs at high 
energies, while the opposite is true at low energies (figure \ref{fig:resultsoverall}, left). 
The differences are more obvious in the zoomed-in plots on figure \ref{fig:resultsoverall}, middle. In this plot, CNNs
outperform BDTs in most cases, despite having higher AUCs. This discrepancy can be understood when looking
at the other zoomed-in curves in figure \ref{fig:resultsoverall}, right.

\begin{table}
\begin{center}
\caption{Summary of the performance (area under curve - AUC) of the different models for each energy band (EB) and for 
each model. }
\begin{tabular}{ llllll }
\br
 
 EB & Num. Evts & AUC BDT & AUC DL18k & AUC DL290k & AUC DL595k \\ 
\mr
 63 to 158 GeV &  354k    & 0.9759 & 0.9600 & 0.9648 & 0.9707 \\  
 100 to 562 GeV &	800k    & 0.9861 & 0.9807 & 0.9851 & 0.9869 \\
 316GeV to 1.8 TeV &  400k  & 0.9904 & 0.9913 & 0.9923 & 0.9920 \\
 1 to 5.6 TeV & 200k & 0.9923 & 0.9950 & 0.9963 & 0.9947 \\
 3.1 to 32 TeV & 120k & 0.9934 & 0.9930 & 0.9958 & 0.9948 \\
\br
\end{tabular}

\label{tab:resultsoverview}
\end{center}
\end{table}

In this plot, it becomes clear that the ROC curve of the BDTs has a steeper start than CNNs, for
all energy bands, which explains the higher overall AUC. Consequently, the best performing approach
will depend on where one applies the cut to separate signal from background. Rejecting as much background
as possible while possibly discarding some signal events would make the BDTs win, while accepting 
as many signal events along with more background would make the CNNs win. Figuring out the most optimal 
cut is always a trade-off between significance and sensitivity, and different cuts might be more adapted 
to specific science goals. We tackled the steepness of the ROC curves problem by training the CNNs with ten times more events. 
In this case it appeared that the overall accuracy does not improve but the slope of the beginning of the ROC 
curve became steeper, in-par with the BDTs. This hints that the current shortcomings 
of CNNs may be addressed simply by augmenting the training datasets with more events. 

\section{Conclusion}
In this paper, we investigated a fair comparison between a state-of-the-art classification technique
for Cherenkov telescopes data and convolutional neural networks. By applying standard CNN
architectures and adapting the Cherenkov data to it, we demonstrated performances that are close
to or already better than existing techniques. This suggest that research in CNNs and other novel 
machine learning approaches should be actively pursued to help achieve the best science output of 
the upcoming CTA observatory.  \\
Many aspects of this investigation will be taken further as there seems to be much room for improvement. 
The datasets could be improved by keeping not only two values but rather the full waveforms. The
neural network architecture could be improved by implementing hexagonal convolutions. The robustness
of the results should be verified by continuing similar comparisons with other, extended datasets.
Finally, the issue of having simulated data that is not identical to the real data should be 
addressed.

\ack
We would like to thank G Maier for kindly providing the list of Monte-Carlo events along with the output of the BDT analysis.
L. Arrabito and J. Bregeon for helping us in our adventure to 
retrieve data from the GRID. The CADMOS (www.cadmos.org) for granting us the compute hours at 
CSCS that made this study possible. The CSCS staff for kindly handling our questions and problems. 
The CTA SAPO for kindly reviewing our paper internally.
We gratefully acknowledge financial support from the agencies and organizations listed here: 
{http://www.cta-observatory.org/consortium\_acknowledgments}

{\linespread{0.8} 
\bibliography{nsb_paper}}

\end{document}